\documentstyle[prc,preprint,tighten,aps,axodraw]{revtex}
\title{\bf Vector Meson Dominance and Rho-Omega Mixing}

\author{S.\ A.\ Coon\thanks{Electronic-mail address: coon@nmsu.edu}}
\address{Physics Department, New Mexico State University, Las Cruces, NM 
 88003}

\author{M.\ D.\ Scadron \thanks{Electronic-mail address:
        scadron@physics.arizona.edu}}
\address{Physics Department, University of Arizona,Tucson, AZ 85721}

\begin{document}

\maketitle
\begin{abstract} 

The scale of a phenomenologically successful charge-symmetry-violating
nucleon-nucleon interaction, that attributed  to meson exchange with a
$\Delta I =1$ rho-omega transition,  is set by  the Coleman-Glashow
$SU(2)$ breaking tadpole mechanism.  A single  tadpole scale has been
obtained from symmetry arguments, electromagnetic meson and baryon
measured mass splittings, and the observed isospin violating  ($\Delta
I=1$) decay $\omega\rightarrow\pi^+\pi^-$. The hadronic realization of
this tadpole mechanism lies in the $I = 1$ $a_0$ scalar meson. We show
that measured hadronic and two-photon widths of the $a_0$ meson, with
the aid of the Vector Meson Dominance  model, recover the
universal Coleman-Glashow tadpole scale.

\end{abstract}

%\vfill

\noindent [Keyword Abstract] charge symmetry breaking, rho-omega mixing, 
nucleon-nucleon potential, $a_0$ scalar meson

\vskip 2truecm
%\vfill
\vskip20pt

\baselineskip 1.0cm               % "double spacing"
\newpage
            
\section{INTRODUCTION}

An isospin-violating effective interaction, with the strength of 
second-order electromagnetic ($em$) theory and labeled $H_{em}$, has long 
been invoked to explain the observed $\Delta I_{z}=1$ meson and baryon 
 diagonal electromagnetic mass 
splittings, and the observed $\Delta I=1$ off-diagonal transitions 
$\rho^\circ-\omega$, $\pi^\circ-\eta$, and $\pi^\circ-\eta'$.  In particular, 
it is the effective $em$ $\rho^\circ-\omega$ transition
\begin{equation}    
	\langle\rho^\circ|H_{em}|\omega\rangle
	\approx -4520 \; {\rm MeV}^2 \ , \label{eqno1}
\end{equation}
as found~\cite{CB} from the observed~\cite{Barkov} isospin violating 
($\Delta I=1$) decay $\omega\rightarrow\pi^+\pi^-$, which underlies the 
dominant charge-symmetry-violating (CSV) nucleon-nucleon interaction of 
refs. \cite{PSC,PSC77}.  The latter CSV NN force is quite successful in 
explaining the observed charge symmetry violation in nuclear physics.  
These observations include NN scattering and bound state (the 
Okamoto-Nolen-Schiffer anomaly) differences in mirror nuclear systems, 
Coulomb displacement energies of isobaric analog states, isospin-mixing 
matrix elements relevant to the 
isospin-forbidden beta decays, and precise measurements of the elastic 
scattering of polarized neutron from polarized 
protons~\cite{Miller,vanoers}.  In addition, the latter CSV 
$\rho^\circ -\omega$ mixing potential is ``natural" ({\em i. e.} 
dimensionless strength coefficients are $\cal{O}$(1) in the contact 
force limit) in the context of low-energy effective Lagrangian approaches 
to nuclear charge symmetry violation~\cite{vkfg}.  In spite of the 
phenomenological success and theoretical plausibility of the CSV 
potential based upon the effective $\Delta I = 1$ Hamiltonian density in 
eq. (1), this potential has been criticized in the recent literature on 
nuclear charge symmetry violation~\cite{qdep}.
 
Alternative approaches~\cite{mixedprop} based, not on data and physical Feynman 
amplitudes, but upon a ``mixed propagator" of field theory, imply a CSV 
NN potential which is neither consistent with the nuclear data nor with 
the naturalness criterion.  One of the misleading conclusions stemming 
from a focus on the mixed propagator (only an ingredient of an NN 
potential) will be discussed in another paper~\cite{CMR}.  In this paper 
we return to the theory behind eq. (1):  the Coleman-Glashow tadpole 
picture~\cite{CG,Coleman} in which both transitions 
$\langle\rho^\circ|H_{em}|\omega\rangle$ and
$\langle \pi^\circ|H_{em}|\eta\rangle$ are given by the tadpole graphs 
of Fig. 1 and the photon exchange graphs of Fig. 2.  We re-examine, in the light of current particle 
data~\cite{PDG}, the numerical accuracy of vector meson dominance 
(VMD)~\cite{VMD}, 
and then use VMD to link measured decays of the $I = 1$ scalar meson $a_0$ to 
the universality of $\Delta I=1$ meson transitions recently 
established~\cite{CS95}.  We close with a discussion of the implications 
of our results for recent conjectures about a direct  $\omega\rightarrow 
2\pi$ coupling~\cite{Kim,MOW,OCTW,Shakin,Tandy}; {\em ie.} a decay not based on 
$\omega\rightarrow\rho\rightarrow2\pi$, which 
is a $G$ parity-violating $\Delta I=1$  transition.

\section{THE PHOTONIC AND TADPOLE COMPONENTS OF $H_{em}$}

The effective $\Delta I=1$ Hamiltonian density $H_{em}$ in (1) was 
originally thought~\cite{CG,Coleman} to be composed of a Coleman-Glashow 
(CG) nonphotonic contact tadpole part (now couched in the language of the
$u_3$ current quark mass matrix $\bar q\lambda^3 q$) along with a photonic
part $H_{JJ}$ involving intermediate photon exchange.  This CG tadpole
mechanism~\cite{CG}
\begin{equation}
	H_{em}= H^3_{tad} + H_{JJ} \,\,\, ,  \label{eqno2}
\end{equation}
with a {\em single} tadpole scale, in fact explains the 13 ground state
pseudoscalar, vector, octet baron,  and decuplet baryon $SU(2)$ observed diagonal
mass splittings without  the introduction of additional free
parameters~\cite{CS95}.
% (and without  use of an additional direct $\Delta I=1$
%$\omega\rightarrow 2\pi$  coupling).

For the off-diagonal $\rho^\circ-\omega$ transition, Gatto~\cite{Gatto} 
first showed that the vector meson dominance (VMD) of Fig. 2 predicts 
the photon exchange contribution
\begin{equation}
	\langle\rho^\circ|H_{JJ}|\omega\rangle = 
	(e/g_{\rho})(e/g_{\omega})m^2_V
  \approx 644\;{\rm MeV}^2\;\;.  \label{eqno3}
\end{equation}
In (\ref{eqno3}) we have used the average $\rho^\circ-\omega$ mass 
$m_V=776$ MeV 
along with the updated VMD ratios $g_{\rho}/e\approx16.6$ and 
$g_{\omega}/e\approx56.3$, with the latter $g_{\rho}$ and $g_{\omega}$ 
couplings found from electron-positron decay rates~\cite{PDG}:
\begin{mathletters}
\begin{equation}
   \Gamma_{\rho ee} = \frac{\alpha^2}{3}m_{\rho}(g_{\rho}^2/4\pi)^{-1}
		\approx 6.77\; {\rm keV}  \;\;,\label{eqno4a}
\end{equation}
\begin{equation}
   \Gamma_{\omega ee} = 
\frac{\alpha^2}{3}m_{\omega}(g_{\omega}^2/4\pi)^{-1}
		\approx 0.60\;  {\rm keV} \;\;, \label{eqno4b}
\end{equation}
\end{mathletters}
leading to $g_{\rho}\approx 5.03$ and $g_{\omega} \approx 17.05$ for 
$e=\sqrt{4\pi\alpha} \approx 0.30282$.  Note that 
Eqs.~(\ref{eqno4a},\ref{eqno4b}) imply the 
ratio $g_{\omega}/g_{\rho} \approx 3.4$, which is reasonably near the 
$SU(3)$ value $g_{\omega}/g_{\rho} = 3$.  Finally, combining the VMD 
$H_{JJ}$ prediction (\ref{eqno3}) with the observed $H_{em}$ transition in 
(\ref{eqno1}), 
one finds the CG tadpole transition using (2) is
\begin{equation}
 \langle\rho^\circ|H^3_{tad}|\omega\rangle \approx -4520 \;{\rm MeV^2}
    -644 \;{\rm MeV^2} \approx -5164 \;{\rm MeV^2}  \;\;. \label{eqno5}
\end{equation}
In fact this off-diagonal CG tadpole scale of (\ref{eqno5}) extracted from 
$\omega\rightarrow 2\pi$ data combined with the VMD scale of 
(\ref{eqno3}) is 
quite close to the CG tadpole scale predicted from the $SU(3)$ diagonal 
vector meson mass splittings~\cite{Gourdin}.  If the $\omega$ is assumed to 
be pure 
nonstrange, the $SU(3)$ prediction becomes
\begin{equation}
\langle\rho^\circ|H_{tad}^3|\omega\rangle=\Delta m^2_{K^*}-\Delta 
m^2_\rho \approx -5120 \; {\rm MeV^2}\;\;, \label{eqno6}
\end{equation}
obtained using the 1996 PDG values, $m_{K^{*+}} \approx 891.6 \; {\rm MeV}$,
$m_{K^{*0}} \approx 896.1 \;{\rm MeV}$ so that $\Delta m^2_{K^*}= 
m_{K^{*+}}^2 - m_{K^{*0}}^2 \approx -8040 \;{\rm MeV^2}$.  While 
$m_{\rho^{+}} \approx 766.9 \; {\rm MeV}$, the more elusive $\rho^0$ mass 
at~\cite{footnote} $m_{\rho^{0}} \approx 768.8 \;{\rm MeV}$ then requires 
$\Delta m^2_{\rho}= m_{\rho^{+}}^2 - m_{\rho^{0}}^2 \approx -2920 \; {\rm 
MeV^2}$.  The difference between $\Delta m^2_{K^*}$ and $\Delta 
m^2_{\rho}$ above then leads to the right hand side of (\ref{eqno6}).

Since only a slight change of $m_{\rho^{0}}$ above can shift $\Delta 
m^2_{K^*}-\Delta m^2_\rho$ by more than 10\%, it is perhaps more 
reliable to exploit $SU(6)$ symmetry 
between the pseudoscalar and vector meson masses,
$ m^2_{K^*} - m^2_{\rho} = m^2_K - m^2_{\pi}$.  But because this SU(6)
relation is valid to within 5\%, it is reasonable to assume the SU(6)
mass difference $ \Delta m^2_{K^*} 
-\Delta  m^2_{\rho} = \Delta m^2_K - \Delta m^2_{\pi}$ also holds.  
Then the $\rho^\circ-\omega$ tadpole transition (\ref{eqno6}) is predicted 
to be~\cite{CS95}
\begin{equation}
	\langle\rho^\circ|H_{tad}^3|\omega\rangle=\Delta m^2_{K^*}-\Delta 
m^2_\rho = \Delta m^2_K - \Delta m^2_{\pi} \approx -5220 \; {\rm MeV}^2 \ ,
	\label{eqno7} 
\end{equation}
because pseudoscalar meson data~\cite{PDG} requires $\Delta m^2_K = 
m_{K^{+}}^2 - m_{K^{0}}^2 \approx -3960 \;{\rm MeV^2}$ and $\Delta 
m^2_{\pi} = m_{\pi^{+}}^2 - m_{\pi^{0}}^2 \approx -1260 \;{\rm MeV}^2$, 
leading to the right hand side of (\ref{eqno7}).  Comparing the similar tadpole 
scales of $\sim -5200 \; {\rm MeV}^2$ in (\ref{eqno5}),(\ref{eqno6}),  and 
(\ref{eqno7}),  we might deduce
from this consistent picture that 
$\langle\rho^\circ|H_{em}|\omega\rangle$ in turn is predicted to have 
the scale $\langle\rho^\circ|H_{em}|\omega\rangle \approx -4500 \;{\rm 
MeV}^2$, as was found from the Barkov $\omega\rightarrow 2\pi$ 
data~\cite{CB,Barkov}.

To emphasize that the above tadpole scale (5)-(7) of the off-diagonal 
$\Delta I =1$ $\rho^\circ-\omega$ transition also holds for the diagonal 
electromagnetic mass differences as well, we briefly review the well-measured 
pseudoscalar $\pi$ and $K$ {\em em} mass splittings.  It has long been 
known~\cite{Das} that $\Delta m^2_{\pi} $ is essentially due to the 
photonic self-interaction mass shifts of the charged and uncharged 
pions~\cite{Tegen}.  As noted in Ref.~\cite{CS95}, this familiar idea takes 
the form in the tadpole picture ($(H_{em})_{\pi^+} = \langle \pi^+ |H_{em}|\pi^+ 
\rangle$, etc):
\begin{mathletters}
\begin{equation}
\Delta m^2_{\pi} \equiv (H_{em})_{\Delta \pi} \equiv (H_{em})_{\pi^+} - 
(H_{em})_{\pi^\circ} = (H_{tad}^3)_{\Delta \pi} + (H_{JJ})_{\Delta \pi} =
(H_{JJ})_{\Delta \pi}\;\;,			\label{eqno8a}
\end{equation} 
where the first equality is due to the CG decomposition (2) and the second 
equality is because $(H_{tad}^3)_{\Delta \pi} = 0$ due to $SU(2)$ symmetry.
However $(H_{tad}^3)_{\Delta K}$ does not vanish in the analogous CG {\em 
kaon} mass splitting relation
\begin{equation}
\Delta m^2_{K} \equiv  (H_{tad}^3)_{\Delta K} + (H_{JJ})_{\Delta K}\;\;.
			\label{eqno8b}
\end{equation}
Then subtracting (8a) from (8b) while using the Dashen PCAC 
observation~\cite{Dashen}
\begin{equation} 
(H_{JJ})_{\pi^\circ} = (H_{JJ})_{K^\circ} = (H_{JJ})_{\bar{K}^\circ}= 0\;\;,
\hspace{.5in} (H_{JJ})_{\pi^+}= (H_{JJ})_{K^+}\;\;,	\label{eqno8c}
\end{equation}
\end{mathletters}
which is strictly valid in the chiral limit, one is led~\cite{CS95} to 
the diagonal pseudoscalar meson tadpole scale
\begin{equation} 
(H_{tad}^3)_{\Delta K} \equiv (H_{tad}^3)_{ K^+} - (H_{tad}^3)_{ K^\circ}
= \Delta m^2_{K} - \Delta m^2_{\pi } \approx -5220 \;\;{\rm MeV}^2\;\;.
			\label{eqno9}
\end{equation}
Extending $\Delta m^2_{K} - \Delta m^2_{\pi }$ to $\Delta m^2_{K^*}-\Delta 
m^2_\rho$ via $SU(6)$ along with $(H_{tad}^3)_{\Delta K} = 
(H_{tad}^3)_{\Delta K^*}$ also by $SU(6)$ symmetry 
yields the diagonal vector meson tadpole scale
\begin{equation}
(H_{tad}^3)_{\Delta K^*} \equiv (H_{tad}^3)_{ K^{*+}} - 
(H_{tad}^3)_{K^{*\circ}} = \Delta m^2_{K^*}-\Delta m^2_\rho \approx -5120 \; 
{\rm MeV^2}\;\;.
                        \label{eqno10}
\end{equation}
Thus we see that the off-diagonal $\rho^\circ - \omega$ tadpole scales (5) 
- (7) together with the diagonal tadpole scales in (9) and (10) are indeed 
universal.  This -5200 MeV$^2$ scale also is approximately valid for 
diagonal baryon masses when one uses quadratic mass formulae for 
baryons~\cite{CS95}.

\section{TADPOLE MECHANISM AND THE $\protect\lowercase{a_0}(980)$}

The $I=1$ $a_0$ scalar meson is assumed to play a unique role in the 
Coleman-Glashow $\Delta I =1$ tadpole mechanism which describes 
$SU(2)$ {\em} mass 
differences and mixing among hadrons~\cite{CG,Coleman,Paver}.  
Furthermore this meson is also almost unique among the scalar mesons 
in that it is
experimentally well established with known decay parameters.  Therefore
one can test the universal tadpole scale of the previous section against
experimental data from an entirely different sector.  We shall see that
this confrontation yields yet another consistent pattern of a $\Delta I
= 1$ universal tadpole scale. 

More 
specifically the $\Delta I = 1$ {\em em} tadpole graphs of Figs. 1 are 
controlled by the $ I = 1$  $a_0(980)$ pole for both 
$a_0\rightarrow\omega\rho^0$ and $a_0\rightarrow\eta\pi^0$ transitions. 
The unknown tadpole $\langle 0|H_{tad}^3|a_0\rangle$ and the $a_0$ 
propagator cancel out of the ratio
\begin{equation}
	\frac{\tilde{F}_{a_0 \rho^0 \omega}}{F_{a_0 \pi^0 \eta_{NS}}} \approx
        \frac{ \langle\rho^\circ|H_{tad}^3|\omega\rangle} 
             {\langle\pi^\circ|H_{tad}^3|\eta_{NS}\rangle} \;\;, \label{eqno11}
\end{equation}
where $\tilde{F}_{a_0 \rho^0 \omega} \equiv m^2_{a_0} F_{a_0 \rho^0 
\omega}$.
The left hand side of (11) can be related to the experimental ratio 
obtained from the PDG rate $\Gamma_{a_0\gamma\gamma}= (0.24\pm 0.08)$ 
keV (assuming $\Gamma_{a_0\pi\eta}$ is overwhelmingly dominant) divided 
by $\Gamma_{a_0\pi\eta}\approx 75$ MeV, midway between the PDG range 
(50-100) MeV:
\begin{equation}
	r = \frac{\Gamma_{a_0\gamma\gamma}}{\Gamma_{a_0\pi\eta}} \approx
       \frac{0.24\; {\rm keV}}{75\; {\rm MeV}} \approx 3.2 \times 
10^{-6}\;\;.  \label{eqno12}
\end{equation} 
This relation is again the vector meson dominance (VMD) model turning 
the vector mesons of Fig 2 and (\ref{eqno11}) into the $\gamma$'s of 
(\ref{eqno12}), since the ``rate" for $a_0 \rightarrow \rho^0 \omega$ cannot 
be directly measured.

To illustrate  
the  VMD model~\cite{VMD} in this context, we first study 
$\omega\pi\gamma$ coupling (times the usual Levi-Civita factor 
$\epsilon_{\mu\nu\alpha\beta} k'^{\mu}k^{\nu} 
\epsilon^{\alpha}\epsilon^{\beta}$) by comparing it to 
$\pi^0\gamma\gamma$ coupling (divided by 2 due to Bose symmetry)
\begin{mathletters}
\begin{equation} 
	F_{\omega\pi^0\gamma} = (g_{\omega}/e) F_{\pi^0\gamma\gamma}/2
          \approx 0.704 \;{\rm GeV}^{-1}    \label{eqno13a}
\end{equation} 
by virtue of VMD turning an $\omega$ into a $\gamma$.  Here 
$F_{\pi^0\gamma\gamma} = \alpha/(\pi f_{\pi}) \approx 0.025 \;{\rm 
GeV}^{-1}$ as found from the axial anomaly or using instead the 
$\pi^0\gamma\gamma$ rate of 7.6 eV~\cite{PDG}.  This VMD prediction 
(\ref{eqno13a})
is in excellent agreement with the measured value~\cite{PDG}
\begin{equation}
        F_{\omega\pi^0\gamma} = \sqrt{12\pi 
\Gamma_{\omega\pi^0\gamma}/k^3 } 
          \approx 0.704 \;{\rm GeV}^{-1}\;\;,    \label{eqno13b}
\end{equation}
where the amplitude $F_{\omega\pi^0\gamma}$ is also weighted by 
$\epsilon_{\mu\nu\alpha\beta} k'^{\mu}k^{\nu} 
\epsilon^{\alpha}\epsilon^{\beta}$.
A similar VMD prediction for $\rho \rightarrow \pi^0 \gamma$ is 
also quite good: The VMD amplitude is 
\begin{equation} 
	F_{\rho\pi^0\gamma} = (g_{\rho}/e) F_{\pi^0\gamma\gamma}/2
          \approx 0.208 \;{\rm GeV}^{-1}    \label{eqno13c}
\end{equation}
using $g_{\rho}\approx 5.03$ found from (4a), while data implies
\begin{equation}
        F_{\rho\pi^0\gamma} = \sqrt{12\pi 
\Gamma_{\rho\pi^0\gamma}/k^3 } 
          \approx 0.222 \;{\rm GeV}^{-1}\;\;,    \label{eqno13d}
\end{equation}
 as extracted from the PDG tables in~\cite{PDG}.
\end{mathletters}

Given this justification of VMD in equs. (13), we follow Bramon and 
Narison~\cite{Bramon} and use VMD to link the CG tadpole mechanism of 
Figs. 2 with 
the observed properties of the $a_0$ meson.  We return to (\ref{eqno11}) 
and note 
that the $a_0 \rho^0 \omega$ coupling in (\ref{eqno11}) is divided by 2 
(as it was 
in (\ref{eqno13a})) when applying VMD to the identical photon transition:
\begin{equation}
	F_{a_0 \rho^0 \omega} \approx (g_{\omega}/e)( g_{\rho}/e )
F_{\pi^0\gamma\gamma}/2  \;\;.   \label{eqno14}
\end{equation}
Since both $F_{a_0 \rho^0 \omega}$ and $F_{a_0\gamma\gamma}$ are 
weighted by the covariant form 
$\epsilon^{*'}_{\mu}\epsilon^{*}_{\nu}(k'\cdot k g^{\mu\nu} - 
k^{\mu}k'^{\nu})$ which squares up to $(k'\cdot k)^2 = 2 
(m^2_{a_0}/2)^2$, the desired amplitude $\tilde{F}_{a_0 \rho^0 \omega} = 
m^2_{a_0} F_{a_0 \rho^0 \omega}$ has the same (GeV)$^1$ mass dimension 
as does $F_{a_0 \pi^0 \eta_{NS}}$.  The latter is given as
\begin{equation}
	F_{a_0 \pi^0 \eta_{NS}} = F_{a_0 \pi^0 \eta} / \cos \phi \approx 
1.35 F_{a_0 \pi^0 \eta}
 \label{eqno15}
\end{equation}
 for the $\eta'-\eta$ mixing angle $\phi \approx 42^{\circ}$ in the 
$NS-S$ quark basis~\cite{angle}.

On the other hand, the theoretical branching ratio rates 
become with VMD~\cite{Bramon}
\begin{equation}
 r = \frac{\Gamma_{a_0\gamma\gamma}}{\Gamma_{a_0\pi\eta}}
  =  \frac{1}{4} \left | \frac{k_{\gamma}}{k_{\eta}}\right | 
     \frac{\tilde{F}^2 _{a_0\gamma\gamma}} {F^2_{a_0\pi\eta}}
  = \frac{4}{4} \left |\frac{k_{\gamma}}{k_{\eta}} \right |
    \left ( \frac{e}{g_{\omega}} \right) ^2
    \left ( \frac{e}{g_{\rho}}\right)^2 \frac{\tilde{F}^2 _{a_0 \rho^0 \omega}}
    {F^2_{a_0\pi\eta}}\;\;\;,   \label{eqno16}
\end{equation}
where $k_{\gamma}=492$ MeV, $k_{\eta} = 321$ MeV, so that 
$|k_{\gamma}/k_{\eta}|\approx 1.53$.
The factor of ${\textstyle \frac{1}{4}}$ in (\ref{eqno16}) corresponds 
to Feynman's rule of ${\textstyle \frac{1}{2}}$ for two identical 
final-state photons, times the numerator factor of ${\textstyle 
\frac{1}{2}}$ in (\ref{eqno16}) coming from the square of the covariant 
factor $\epsilon^{*'}_{\mu}\epsilon^{*}_{\nu}(k'\cdot k g^{\mu\nu} -
k^{\mu}k'^{\nu})$ (times $m^2_{a_0}$ which is absorbed into the 
definition of $\tilde{F} _{a_0\gamma\gamma}$). Finally a factor of 4 
in the numerator of the right hand side of (\ref{eqno16}) is due to the 
square of the VMD relation (\ref{eqno14}).

Substituting the observed $r$ from (\ref{eqno12}) back into the 
theoretical ratio (\ref{eqno16}) and converting to the $\eta_{NS}$ 
basis using (\ref{eqno15}) then leads to the amplitude ratio
\begin{mathletters}
\begin{equation}
\frac{\tilde{F}_{a_0 \rho^0 \omega}}{F_{a_0 \pi^0 \eta_{NS}}} \approx 
1.0\;\;\;,
 \label{eqno17a}
\end{equation}
a result which stems only from observed properties of the $a_0$ meson 
and VMD.

If one goes further and identifies the transition 
$\langle 0|H_{em}|a_0\rangle$ as the Coleman-Glashow tadpole, the 
VMD-phenomenological estimate of unity for the ratio (\ref{eqno17a}) 
requires the tadpole ratio in (\ref{eqno11}) and in Figs. 1 also to be 
unity
\begin{equation}
\frac{ \langle\rho^\circ|H_{tad}^3|\omega\rangle}
             {\langle\pi^\circ|H_{tad}^3|\eta_{NS}\rangle} \approx 
1.0\;\;\;,
 \label{eqno17b}
\end{equation}
\end{mathletters}
This hadronic picture of the CG tadpole is consistent 
with the universal $SU(6)$ tadpole scale already 
obtained in Ref. \cite{CS95} and reviewed in Section 2.

Implicit in this identification is the conventional
$\bar{q}q$ assignment of the $a_0$.  According to ref.~\cite{Bramon},
the tadpole mechanism fails to predict the experimental
$\Gamma_{a_0\gamma\gamma}$ if the $a_0$ is considered to be a
$\bar{q}qq\bar{q}$ state.  Recent $K$-matrix analyses of meson partial 
waves from a variety of three-meson final states obtained from $\pi^- p$ 
and $p\bar{p}$ reactions show rather convincingly that both the $I=1$ 
$a_0(980)$ and 
$I=0$  $f_0(980)$ mesons are formed from the bare states which are members of 
the lowest $\bar{q}q$ nonet~\cite{Anisovich}.  Recent theoretical 
developments supporting the assignment of the $a_0$ to   the scalar
meson $\bar{q}q$ nonet are reviewed in Ref.~\cite{Napsuciale}.

The modern identification of the tadpole scale with the mass 
difference of the up and down current quarks:
\begin{equation}
	H^3_{tad} = {\textstyle \frac{1}{2}}(m_u - m_d)\bar{q}\lambda_3 q,
\label{eqno18}
\end{equation}
suggests a parallel treatment of the hadronic and two-photon couplings 
of the $a_0$ based on the three point function method in QCD sum rules.
One begins with VMD to express another measured ratio
\begin{equation}
	\frac{\Gamma_{a_0\gamma\gamma}}{\Gamma_{\pi^0\gamma\gamma}}
	\approx \frac{m_{a_0}^3}{m_{\pi}^3} \frac{g_{a_0 \rho^0 \omega}}
	{g_{\pi^0 \rho^0 \omega}}
\label{eqno19}
\end{equation}
in terms of the strong coupling constants $g_{a_0 \rho^0 \omega}$ and 
$g_{\pi^0 \rho^0 \omega}$.  The latter coupling constant ratio is then
estimated with the aid of QCD sum rules which bring in the up and down
current quark masses. The result is again a reasonable value for
$\Gamma_{a_0\gamma\gamma}$ and the ratio $r$ of (\ref{eqno12}).  This
QCD sum rule treatment of the $a_0$ decays is consistent with, but does not
really give new information about the Coleman-Glashow tadpole.  So we do
not describe it further and refer to Refs. \cite{Bramon,Dominguez} for
a detailed account of this QCD sum rule program.

\section{DISCUSSION}

In Section 3, vector meson dominance was used to  link measured decays
of the $I = 1$ scalar meson $a_0$ to  the universality of $\Delta I=1$
meson transitions $\langle\rho^\circ|H_{em}|\omega\rangle$ and $\langle
\pi^\circ|H_{em}|\eta\rangle$ recently  established~\cite{CS95}.  One
element of this universality is the value of the effective $em$
$\rho^\circ-\omega$ transition found~\cite{CB} from the
observed~\cite{Barkov} isospin violating ($\Delta I=1$) decay
$\omega\rightarrow\pi^+\pi^-$. To obtain this value, the decay is
analyzed as $\omega\rightarrow\rho\rightarrow2\pi$~\cite{PSC77}.  It
has  recently been asserted~\cite{Kim}  that a   direct
$\omega\rightarrow 2\pi$ coupling is not only necessary on general
principles, but  that a significant coupling is supported by a
theoretical QCD sum rule analysis of a isospin-breaking correlator of
vector currents.  This assertion has prompted  reanalyses of the data
on  $e^+ e^-\rightarrow\pi^+ \pi^-$~\cite{MOW,OCTW}, and the modeling
of this putative contact  $\omega\rightarrow 2\pi$ term in two quark based
models of $\rho^{\circ}-\omega$ mixing~\cite{Shakin,Tandy}. The results
of these three investigations are somewhat mixed. Here they are given  
as the ratio $G = g_{\omega_I\pi\pi}/g_{\rho_I\pi\pi}$, where 
$\omega_I$ and $\rho_I$ are the basis states of pure isospin $I=0$ and
$I=1$, respectively.  The   data analysis of Ref.\cite{OCTW} suggests
$G\approx 0.10$, the coupled Dyson-Schwinger equations approach~\cite{Tandy}
finds a value five times smaller ($G\approx 0.017$), and the
generalized Nambu-Jona-Lasinio model~\cite{Shakin}  predicts a value which is a
further factor of four smaller ($G\approx 0.004$). The last very small
ratio would make direct $\omega$ decay a relatively unimportant
contribution to the calculation of $\rho^{\circ}-\omega$ mixing from
the data. On the other hand, the isospin breaking from direct $\omega$
decay suggested by the data analysis \cite{OCTW} is a huge 10\% rather than the few 
percent usually found for isospin breaking (cf. the  2\% Coleman-Glashow
ratio reviewed in the Appendix of Ref.\cite{CS95}).  This in turn
drives the value of $\langle\rho^\circ|H_{em}|\omega\rangle$ up to
$\approx -6830$ MeV$^2$,  rather far from the value of $\approx -4520$
MeV$^2$ (quoted in Eq. (1)) obtained from the same data when this
putative contact term is ignored.  It is the latter value which was
shown in sections 2 and 3 to be consistent with the universality
discussed there.  The most recent extraction of a $\rho^{\circ}-\omega$
mixing parameter from the data eschews such a separation into a contact
$\omega\rightarrow\pi^+\pi^-$ and mixing
$\omega\rightarrow\rho\rightarrow2\pi$ term on the grounds that it is 
model dependent~\cite{Gardner,width}.
As we have seen, a significant contact G-parity violating $\omega\rightarrow
2\pi$ coupling  would increase in magnitude the value of 
$\langle\rho^\circ|H_{em}|\omega\rangle$ to such an extent that it would not be
consistent with the off-diagonal  $\rho^\circ - \omega$ tadpole scales (6)  and
(7), nor with  the diagonal tadpole scales in (9) and (10), nor with the
diagonal tadpole scales obtained from the baryon mass differences~\cite{CS95}.
Furthermore, such a large value of $\langle\rho^\circ|H_{em}|\omega\rangle$ is
not consistent with the $\Delta I = 1$ universal tadpole scale obtained in (17)
with the aid of the Vector Meson Dominance  (VMD) model.  In view of the
inconsistency with the global Coleman-Glashow picture and the limited support
from quark based models~\cite{Shakin,Tandy} for a contact $\omega\rightarrow
2\pi$ which violates G-parity, it is instructive to look at the analogue
four-point contact term in the {\em strong} decay $\omega\rightarrow 3\pi$. 
This term was introduced on general grounds some time after the
suggestion~\cite{GMSW} that  $\omega\rightarrow\rho\pi\rightarrow 3\pi$ will
dominate the  $\omega\rightarrow 3\pi$ transition.  This dominance of this VMD
$\omega\rho\pi$ pole diagram model  has been confirmed by the experimental
study of the   $e^+ e^-\rightarrow3\pi$ reaction~\cite{exp}.  In fact, a
contact term large enough to satisfy a low-energy theorem~\cite{VVA} in the
pseudoscalar sector(the AVV anomaly)  spoils agreement with this data. The
history of the fate of the contact $\omega\rightarrow 3\pi$ term can be traced
in Ref. \cite{Silagadze}, which concludes that ``nowadays the existence and
magnitude of the contact term can be extracted neither from theory, nor
experiment."

We suggest that a similar fate may be store for the proposed G-parity violating
$\omega\rightarrow 2\pi$ contact term.  While its effects cannot be cleanly
isolated from  data~\cite{MOW,OCTW}, in contrast to the proposed strong
interaction $\omega\rightarrow 3\pi$ contact term, nevertheless the
introduction of this  G-parity violating $\omega\rightarrow 2\pi$
contact term  is inconsistent with  the $SU(6)$ prediction (7), the universal
CG tadpole scale~\cite{CS95}, and, as shown in section 3, the measured decay
properties of the $a_0$ meson.

\pagebreak

\vspace{72pt}

\begin{picture}(300,110)(0,0)
\DashArrowLine(0,25)(50,25){5}
\DashArrowLine(50,25)(100,25){7}
\DashArrowLine(50,25)(50,75){6}
\CCirc(50,90){15}{Red}{White}
\Text(50,90)[]{$H_{tad}^3$}
\Text(25,15)[]{$\rho^{\circ}$}
\Text(75,15)[]{$\omega$}
\Text(60,50)[]{$a_{\circ}$}
\DashArrowLine(200,25)(250,25){5}
\DashArrowLine(250,25)(300,25){7}
\DashArrowLine(250,25)(250,75){6}
\CCirc(250,90){15}{Red}{White}
\Text(250,90)[]{$H_{tad}^3$} 
\Text(225,15)[]{$\pi^{\circ}$}
\Text(275,15)[]{$\eta_{NS}$}
\Text(260,50)[]{$a_{\circ}$}
\end{picture}

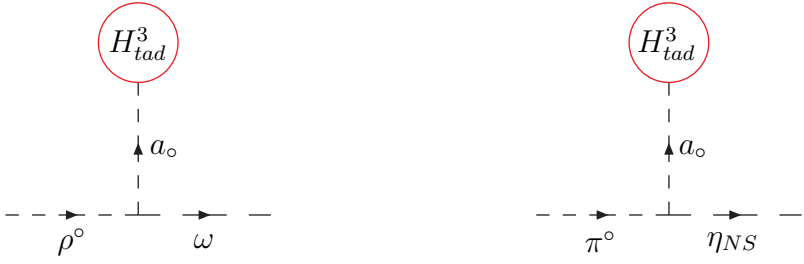
Figure 1 $a_{\circ}$ meson tadpole diagrams for  the CSB $\Delta I = 1$ 
transitions 
$\langle\rho^\circ|H_{tad}^3|\omega\rangle$
and 
$\langle\pi^\circ|H_{tad}^3|\eta_{NS}\rangle$.  According to Coleman
and Glashow~\cite{CG}
, these are diagrams that can be broken into two parts,
connected only by the scalar meson $a_0$ line, such that one part is
the scalar tadpole  $\langle 0|H_{tad}^3|a_0\rangle$, represented by
the circle, and the other part involves only the $SU(3)$ invariant
strong interactions.  The latter interactions, in this case,  
$a_0\rightarrow\omega\rho^0$ and $a_0\rightarrow\eta\pi^0$ transitions,
are represented by the coupling constants of (11).

\vspace{72pt}

\begin{picture}(300,100)(0,0)
\SetColor{Red}
\Photon(100,50)(200,50){5}{10}
\SetColor{Black}
\Vertex(100,50){2}
\Vertex(200,50){2}
\DashArrowLine(0,50)(100,50){5}
\DashArrowLine(200,50)(300,50){7}
\Text(50,40)[]{$\rho^{\circ}$}
\Text(250,40)[]{$\omega$}
\Text(150,40)[]{\Red{$\gamma$}}
\end{picture}\\

Figure 2  The current-current contribution 
$\langle\rho^\circ|H_{JJ}|\omega\rangle$  to 
$\langle\rho^\circ|H_{em}|\omega\rangle$.

\end{document}